\documentclass{pasj00}

\def\tsutsui{$E_{\rm p}$--$T_{\rm L}$--$L_{\rm p}$ }
\def\yonetoku{$E_{\rm p}$--$L_{\rm p}$ }
\def\amati{$E_{\rm p}$--$E_{\rm iso}$ }

\begin{document}
\SetRunningHead{R. Tsutsui et al.}{Classification of long gamma-ray bursts}
\title{Identifying Subclasses of Long Gamma-Ray Bursts with  Cumulative
Light Curve Morphology of Prompt Emissions}

\author{Ryo
\textsc{Tsutsui},\altaffilmark{1}\email{tsutsui@resceu.s.u-tokyo.ac.jp}
Takashi \textsc{Nakamura},\altaffilmark{2}
Daisuke \textsc{Yonetoku},\altaffilmark{3}
Keitaro \textsc{Takahashi},\altaffilmark{4}
and
Yoshiyuki \textsc{Morihara}\altaffilmark{3}
}
\altaffiltext{1}{Research Center for the Early Universe, School of
Science, University of Tokyo, Bunkyo-ku, Tokyo 113-0033, Japan}
\altaffiltext{2}{Department of Physics, Kyoto University,
Kyoto 606-8502, Japan}
\altaffiltext{3}{Department of Physics, Kanazawa University, Kakuma,
Kanazawa,
Ishikawa 920-1192, Japan}
\altaffiltext{4}{Faculty of Science, Kumamoto University,
Kurokami, Kumamoto, 860-8555, Japan}


%


\maketitle

\begin{abstract}

We argue a new classification scheme of long gamma-ray bursts (LGRBs)
using the morphology of the cumulative light curve of the prompt emission.
We parametrize the morphology by the absolute deviation from their constant
luminosity ($ADCL$) and derive the value for 36 LGRBs which have
spectropic redshifts,
spectral parameters determined by the Band model, 1-second peak fluxes,
fluences,
and 64-msec resolution light curves whose peak counts are 10 times
larger than
background fluctuations. Then we devide the sample according to the
value of ADCL
into two groups ($ADCL < 0.17$ and $ADCL > 0.17$) and, for each group,
derive the spectral peak energy $E_{\rm p}$ -  peak luminosity $L_{\rm p}$
correlation and the Fundamental Plane of LGRBs, which is a correlation
between the spectral peak energy $E_{\rm p}$, the luminosity time
$T_{\rm L}$
($\equiv E_{\rm iso}/L_{\rm p}$ where $E_{\rm iso}$ is isotropic energy)
and the peak luminosity $L_{\rm p}$.
We find that both of the correlations for both groups are statistically
more significant compared with ones derived from all samples.
The Fundamental Planes with  small and large ADCL are given by
$L_{\rm p}=10^{52.53\pm 0.01}(E_{\rm p}/10^{2.71}{\rm keV})^{1.84\pm 0.03}
(T_{\rm L}/10^{0.86}{\rm sec})^{0.29\pm0.08}$
with $\chi^2_{\nu}=10.93/14$ and
$L_{\rm p}=10^{52.98\pm0.08}(E_{\rm p}/10^{2.71}{\rm keV})^{1.82\pm 0.09}
(T_{\rm L}/10^{0.86}{\rm sec})^{0.85\pm 0.27}$
with $\chi^2_{\nu}=7.58/8$, respectively. This fact implies the existence
of subclasses of LGRBs characterized by the value of $ADCL$.
Also there is a hint for the existence of the intermediate-$ADCL$ class
which deviates from both fundamental planes. Both relations are
so tight that our result provides a new accurate distance measurement
scheme up to the high redshift universe.

\end{abstract}


\section{Introduction}

In spite of the discovery of the observational diversity of the afterglow,
parent galaxy, and environment properties of gamma-ray bursts (GRBs)
in {\it Swift} era, the prompt emission studies over the last 10 years
have not succeeded in revealing telltale GRB subclasses beyond
the short-hard and long-soft classification \citep{Kouveliotou:1993}.
Except for some unusually low luminous events which are often associated
with Type Ic Supernovae (e.g., 980425, 060218), all long gamma-ray bursts
(LGRBs) seem to have similar properties and no subclasses.

In \citet{Tsutsui:2009b}, we found a correlation between the spectral peak
energy $E_{\rm p}$, the luminosity time $T_{\rm L}$ and the peak
luminosity $L_{\rm p}$
(Fundamental Plane\footnote{{\it The Fundamental Plane} is a two-dimensional
family first discovered between velocity dispersion, mean surface brightness
and luminosity of elliptical galaxy \citep{Djorgovski:1987}. The term is
used for various two-dimensional families of other objects, such as galaxy
clusters \citep{Schaeffer:1993}, black hole activity \citep{Merloni:2003}
and X-ray gas in normal galaxies \citep{Diehl:2005}.} of LGRBs).
However, the correlation has a relatively large number of outliers
($\sim 30\%$) and they deteriorate the usefulness of LGRBs as standard
candles which probe the expansion history of universe \citep{Tsutsui:2011}.
The outliers may also imply the existence of subclasses and if we can
identify them it would
not only reduce the ambiguity of the correlation but also give some
information
on the origin of LGRBs.

In this paper, we show the statistical significance of the correlation
improves by considering the morphology of the cumulative light curve of
the prompt emission. Specifically, first, we define {\it the absolute
deviation from their constant luminosity ($ADCL$)} of the cumulative light
curve. Then we derive the correlation both for small and large $ADCL$
groups and they are shown to be tighter than one derived by all LGRBs.
We also find that dimmer outliers of \citet{Tsutsui:2011} have intermediate
$ADCL$ values.

\section{Data Description}
\label{sec:data}
Since the discovery of a correlation between the peak energy $E_{\rm p}$ and
the isotropic energy $E_{\rm iso}$ \citep{Amati:2002}, several correlations
between $E_{\rm p}$ and other brightness
\citep{Yonetoku:2004,Ghirlanda:2004a}, were proposed. There are also some
studies to improve these correlations by adding third parameters of GRBs
\citep{Liang:2005,Firmani:2006, Tsutsui:2009b,Yu:2009}.
These correlations are thought to be important not only to investigate
the nature of their emission mechanism but also to measure distance up to
the high redshift universe where no other distance indicators have been
observed
\citep{Ghirlanda:2004b,Schaefer:2007,Kodama:2008,Liang:2008,Cardone:2009,Tsutsui:2009b}.

In spite of their attraction, there have been many debate against these
correlations
\citep{Nakar:2005,Band:2005,Butler:2007,Rossi:2008,Collazzi:2008,Shahmoradi:2011}.

Most of these discussions, however, are based on incomplete data without
redshifts, and data with many systematic errors which conceal the
nature of GRBs.
Therefore, we do not take these works into account in this paper.

We have investigated origins of systematic errors of the spectral -
brightness correlations
\citep{Yonetoku:2010,Tsutsui:2010}, and succeeded to remove them
\citep{Tsutsui:2011}.
As a result, we have found that the $E_{\rm p}$--$T_{\rm L}$--$L_{\rm
p}$ correlation possibly divides LGRBs into at least three subclasses.
In this paper,  we study a new classification scheme based on a relationship
between the residual from the $E_{\rm p}$--$T_{\rm L}$--$L_{\rm p}$
correlation and long term pulse shape within the whole duration of bursts.

Before we study the classification of LGRBs, here we update and recompile
the database of \citet{Yonetoku:2010} and \citet{Tsutsui:2011}.
There are more than 120 GRBs with known redshifts and spectral
parameters of time-integrated spectrum up to October 2011.
However more than half of their spectral parameters determined by not
the Band model \citep{Band:1993}, but by the Cut-off power law model
\citep{Pendleton:1997} because of the lack of number of high energy
photon and/or
the limited energy range of detectors, which tends to overestimate
spectral peak energies
\citep{Kaneko:2006,Krimm:2009,Tsutsui:2010}.
Using these biased data, the difference between subclasses is smoothed out
with their systematic errors \citep{Tsutsui:2011}.
Therefore we do not deal with these data from the beginning in this paper.
Then, we use only 44 LGRBs with known redshifts, spectral parameters
determined by the Band model, observer frame 1-second peak fluxes,
fluences, and 64 msec resolution light curves.

For the GRBs detected by GBM/{\it Fermi}, the spectral analysis is
performed with the software package
RMFIT\footnote{http://fermi.gsfc.nasa.gov/ssc/data/analysis/} (version
3.3rc8) and the GBM Response Matrices v1.8.
We analyze the time-integrated spectrum using CSPEC data, from 8 keV
to 40 MeV with 1.024 sec temporal resolution, following the guidance of
the RMFIT
tutorial\footnote{http://fermi.gsfc.nasa.gov/ssc/data/analysis/user/vc\_rmfittutorial.pdf}.

For other GRBs, there are spectral parameters of individual GRBs
reported by different observation teams and different authors,
so that we choose the data fitted by the Band model with the smallest
uncertainties.
Because we did not know the systematic difference between $E_{\rm p}$
estimated with the Band model and $E_{\rm p}$ with the CPL model in our
previous studies, the data in this paper is not exactly the same as the
data in \citet{Yonetoku:2010} and \citet{Tsutsui:2011}.

The KONUS team often reports not 1 sec peak fluxes but 64,
128, 256 msec fluxes, so that we convert these peak fluxes to 1 sec peak
fluxes
with 64 msec light curves obtained from the archive of all the GCN/KONUS
Notices and Light
Curves\footnote{http://gcn.gsfc.nasa.gov/konus\_grbs.html}
\citep{Yonetoku:2010}.

For all data reported with errors at 90\% confidence level, we convert
it to errors at 68\% one by multiplying $1/1.645$.
In table \ref{tab2}, we summarize redshifts, spectral parameters, observed
1-second fluxes and fluences with detector's energy band and
their references of 44 LGRBs.

From these parameters, bolometric 1-second luminosities ($L_{\rm
p,1s}^{\rm obs}$)
and isotropic equivalent energies ($E_{\rm iso}$) between 1-10,000 keV
in GRB rest frame
are calculated by following equations:
\begin{eqnarray}
\label{eq:conversion}
L_{\rm p,1}^{\rm obs} &=& 4\pi d_{\rm L}^2 P_{\rm p,obs} \times
\frac{\int_{1/(1+z)}^{10,000/(1+z)} E \times N(E) dE}
    {\int_{E_{\rm min}}^{E_{\rm max}} N(E) dE}\\
~~~&&\hspace{4cm}{\rm    (erg~s^{-1})},\nonumber \\
\label{eq:conversion2}
E_{\rm iso} &=& \frac{4\pi d_{\rm L}^2}{1+z} S_{\rm obs} \times
\frac{\int_{1/(1+z)}^{10,000/(1+z)} E \times N(E) dE}
{\int_{E_{\rm min}}^{E_{\rm max}} E \times N(E) dE}\\
~~~&&\hspace{4cm}{\rm (erg)}.\nonumber
\end{eqnarray}
where $N(E)$ is the Band model in units of ${\rm photons~cm^{-2}
s^{-1} keV^{-1}}$ \citep{Band:1993} and $d_{\rm L}$
is the luminosity distance in units of ${\rm cm}$.
We should note that $L_{\rm p,1}^{\rm obs}$'s are
calculated within 1-second in observer frame, which must be converted to
rest $\tau$-second peak luminosities $L_{\rm p,\tau}^{\rm rest}$'s
\citep{Yonetoku:2010,Tsutsui:2011}.
In this paper, we adopt $\tau=2.752$ second for all GRBs. 
\ref{sec:FP}).

There is another parameter characterizing a time scale of
GRBs. \citet{Firmani:2006} used the high signal time $T_{0.45}$ to improve
the $E_{\rm p}$--$L_{\rm p}$ correlation.
Here we adopt another time scale, the luminosity time $T_{\rm L}$, first
used by \citet{Tsutsui:2009b} defined as,
\begin{equation}
T_{\rm L}\equiv \frac{E_{\rm iso}}{L_{\rm p}},
\end{equation}
where an error of $T_{\rm L}$ is calculated from the error propagation
formula without the cross-term.

In this paper, we analyze light curves of LGRBs to identify subclasses
of LGRBs.
For secure analysis of light curves we use only the bursts whose
background fluctuation
-  peak counts ratio are less than 0.1.
This reduces the number of GRBs from 44 to 36.
Then we analyze 36 bursts in following sections.
In table \ref{tab:intrinsic}, we summarize the intrinsic properties of
the 36 GRBs.


\section{Fundamental Plane of Long Gamma-Ray Bursts}
\label{sec:FP}
The time scales which improve the $E_{\rm p}$--$L_{\rm p}$ correlation
have been studied in the past.
Some obtained positive results \citep{Firmani:2006,Tsutsui:2009b} and
the others negative results \citep{Rossi:2008,Collazzi:2008}.
There are several reasons that the previous studies could not establish
the tight correlation between the spectral peak energy, the time scale, and
the peak luminosity : systematic errors of peak energies estimated by the
CPL model, systematic errors of peak luminosities estimated in 1 seconds
of observer, and contaminations of short GRBs with extended emission or
other outliers. They have all disturbed the estimation of the best fit
function and their dispersion so that such tight correlations have never
been
reproduced without the studies of the systematics as mentioned above
\citep{Yonetoku:2010,Tsutsui:2010,Tsutsui:2011}.

Here we briefly describe the method developed in \citet{Tsutsui:2011}.
We assume  a linear correlation between $E_{\rm p}$, $T_{\rm L}$,
and $L_{\rm p}$ in a logarithmic form as:
\begin{eqnarray}
\log L_{\rm p}(E_{\rm p},T_{\rm L})= A
&+& B\log \left(\frac{E_{\rm p}}{10^{2.71}{\rm keV}}\right) \nonumber\\
&& + C\log \left(\frac{T_{\rm L}}{10^{0.86}{\rm sec}}\right),
\end{eqnarray}
where $A,~B$, and $C$ are free parameters of the model. If we adopt $C=0$,
and $C=-1$, this is identical with the \yonetoku and \amati
correlations, respectively.

The corresponding $\chi^2$ function is given by :
\begin{eqnarray}
&&\chi^2(A, B, C, \tau)=\sum_{i}z_i^2, \\
&&z_i=\frac{\log L_{{\rm p},\tau i}^{\rm rest}-\log
			     L_{\rm p}(E_{{\rm p},i},T_{{\rm L},i})}
{\sqrt{(1+2C)\sigma_{\log L_{\rm p},i}^2+B^2\sigma_{\log E_{\rm
p},i}^2+C^2\sigma_{\log T_{\rm
L},i}^2+\sigma_{\rm int}^2}}
\end{eqnarray}
The factor $2C$ in front of $\sigma_{\log L_{\rm p}}$ comes from the fact
that the definition of $T_{\rm L}$ includes $L_{\rm p}$.

Because the $\chi^2$ value strongly depends on the existence of outliers
which deviate from the Gaussian distribution, we first optimize the
Lorentzian merit function, given by :
\begin{eqnarray}
M(A, B, C, \tau)=\sum_{i}\ln (1+\frac{1}{2}z_i^2),
\end{eqnarray}
with $\sigma_{\rm int}=0$.
Because the Lorentzian merit function is less affected by
outliers which do not follow Gaussian distribution, we use the
parameters optimizing the Lorentzian merit as the tentative values to
eliminate outliers and to estimate robust $\sigma_{\rm int}$ value.

To estimate robust $\sigma_{\rm int}$, we use the robust standard
deviation given by,
\begin{equation}
\sigma_{\rm RSD}
\equiv
\frac{{\rm median}
      \left[ |\log L_{{\rm p},\tau i}^{\rm rest}
              - \log L_{\rm p}(E_{{\rm p},i},T_{{\rm L},i})|
      \right]}
    {0.6745}.
\label{eq:RSD}
\end{equation}
and then we obtain robust $\sigma_{\rm int}$ as follows :
\begin{eqnarray}
&&\sigma_{\rm int}^2
= \sigma_{RSD}^2\nonumber \\
&&  - \frac{1}{N}
   \sum_{i}^{N}
   \left\{ (1 + 2 C) \sigma_{\log L_{{\rm p},i}}^2 + B^2 \sigma_{\log
    E_{{\rm p},i}}^2
           + C^2 \sigma_{\log T_{{\rm L},i}}^2 \right\}.\nonumber\\
\label{eq:int}
\end{eqnarray}

Now that we have tentative sets of parameters and $\sigma_{\rm int}$, we
can compute, for each sample, $t=|z_i|$ and the two-tailed P-value from
the $t$ distribution with $(N-3)$ degrees of freedom.
We adopt a threshold probability as $P_{\rm th}=Q/N$ for all GRBs, and the
GRBs which have P-value smaller than $P_{\rm th}$ are regarded as
outliers and
eliminated from the following chi square analysis.
Here Q is arbitrary number less than 1.
If it is not mentioned we always use $Q=0.2$ in this paper.
Therefore we now have data sets without outliers and $\sigma_{\rm int}$, and
we can perform the ordinary chi square analysis to estimate the best fit
parameters and their uncertainties.

\begin{figure}[ht]
\begin{center}
\FigureFile(85mm,85mm){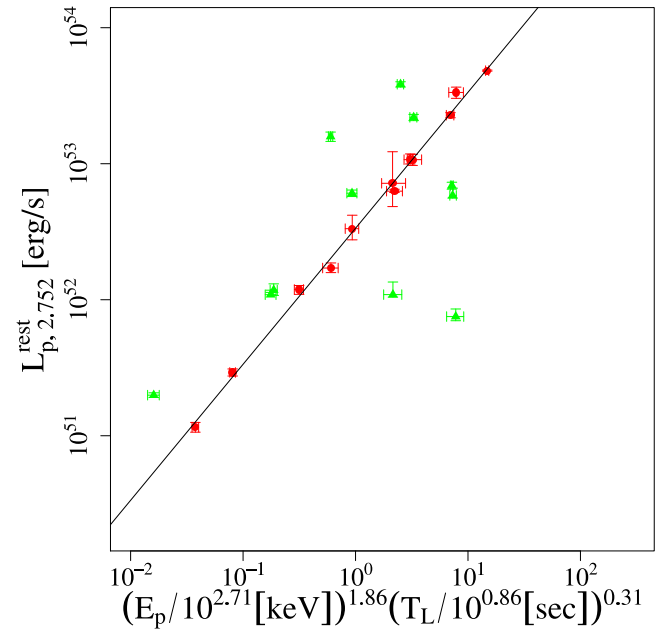}
\end{center}
\caption{The \tsutsui diagram of updated platinum sample.
The green triangles represent outliers.
}
\label{fig:allsample}
\end{figure}

First we apply this method to our updated sample, both the whole sample
and the platinum sample with $\sigma_{\rm E_{\rm p}}/E_{\rm p}< 0.1$.
In \citet{Tsutsui:2011}, using only the platinum sample we obtained the
very tight correlation with six outliers.
The reason for only the platinum sample is to distinguish possible
outliers or subclasses from
the observational errors. As a result in \citet{Tsutsui:2011} we
suggested that there might be three  subclasses in LGRBs.
In figure \ref{fig:allsample}, we plot the \tsutsui diagram obtained  by
the present platinum sample.
The platinum events consistent with the \tsutsui correlation, that is,
on plane events, are marked with red circles,
and  outliers with green triangles.
In table \ref{tab:allsample}, we summarize  the 	best fit parameters,
fraction of outliers ($N_{\rm out}/N$),
intrinsic dispersion ($\sigma_{\rm int}$) and reduced chi square
($\chi_{\nu}^2$) both for the whole sample and the platinum sample.
As table 1 shows, the best fit  values of parameters, $\sigma_{\rm
int}$, and the fraction of outliers ($N_{\rm out}/N$)
depend on  the sample. We see that $\sigma_{\rm int}$ for the whole
sample is 2.6 times larger than that for the platinum sample.
While  the outliers exist for both samples so that the distributions are
not Gaussian for both samples.
This fact suggests that strong non-gaussianity, perhaps bimodality of
the scatters around the best fit function exists.
As figure \ref{fig:allsample} indicates, there might be a subclass of
GRBs (green triangles in the upper left corner)
which might be another correlation nearly parallel to the \tsutsui
correlation. This situation is similar to
PoPI and  PoPII Cepheid variable correlations that are parallel due to
the difference of metallicity.
In practice, however, it is sometime difficult to remove all the
outliers as shown in Monte Carlo simulations
in Appendix A in Tsutsui et al. (2011)
even if we use the outlier detection technique based on robust statistics .
This might be the reason for the best fit parameters and $\sigma_{\rm
int}$ for platinum sample in table 1 being
different from the ones in Tsutsui et al. (2011) although the tightness
of the correlation is unchanged.
This fact suggests that there is a large fraction of outliers close to
the best fit function.
If we can remove these outliers using other observational properties in
advance,
the Fundamental Plane of LGRBs will be more statistically significant
and robust. The purpose of this paper is to find what is
this other observational property like the metallicity in PoPI and
PoPII Cepheid variable correlations.

In the following sections, we introduce a new parameter of LGRBs,
{\it the absolute deviation from their constant luminosity ($ADCL$)}.
We show that dividing LGRBs according to $ADCL$-value significantly
improves both of the \yonetoku correlation and
the Fundamental Plane of GRBs.
That is, we show that ADCL is a new observational property what we have
been seeking for.

\begin{table}[htdp]
\caption{Fitting results for our new sample.}
\begin{center}
\begin{tabular}{|c|c|c|c|c|}
\hline
sample &  best fit  & $\sigma_{\rm int}$ & $N_{\rm out}$/$N$ &
$\chi^2_{\nu}$ \\ \hline
all & $(52.63, 1.63, -0.08)$  & 0.13 & 8/36 & 20.9/26 \\ \hline
platinum & $(52.52, 1.86, 0.31)$ 	&  0.05	& 11/23 & 3.81/10 \\ \hline
\end{tabular}
\end{center}
\label{tab:allsample}
\end{table}

\section{Absolute Deviation from Constant Luminosity (ADCL)}
\label{sec:ADCL}
To quantify the absolute deviation from their constant luminosity
($ADCL$), we analyze 64 msec
resolution light curves of BATSE, KONUS, Swift, and GBM.
We summarize the energy band of each detector in table \ref{tab:energyband}.
\begin{table}[htdp]
\caption{The energy band of light curves for each detector}
\begin{center}
\begin{tabular}{|c|c|c|c|}\hline
BATSE & KONUS & Swift & GBM\\ \hline
20-1000 keV & 50-200 keV & 15-150 keV & 8-1000 keV \\ \hline
\end{tabular}
\end{center}
\label{tab:energyband}
\end{table}%

We determine the background intervals before and after bursts for BATSE,
Swift, and GBM light curves and make linear fit of background using
least square method.
For KONUS light curves, there are no pubic data before bursts so that
we use the background intervals only after bursts.
Then we subtract the backgrounds from 64 msec light curves
interpolating or extrapolating the fit across the whole duration of bursts.
From these background subtracted light curves, we determine the duration
of bursts by ourselves.
The most popular duration of GRBs is $T_{90}$ defined by
\citet{Kouveliotou:1993}.
$T_{90}$ is the time during which the cumulative counts increase from
5\% to 95\%.
However, we find that $T_{90}$ tends to lose weak long tails of bursts
which are essential to classify the bursts.
Thus, we use $T_{98}$ to define $ADCL$ instead of $T_{90}$.

\begin{figure}[htbp]
\begin{center}
\FigureFile(85mm,85mm){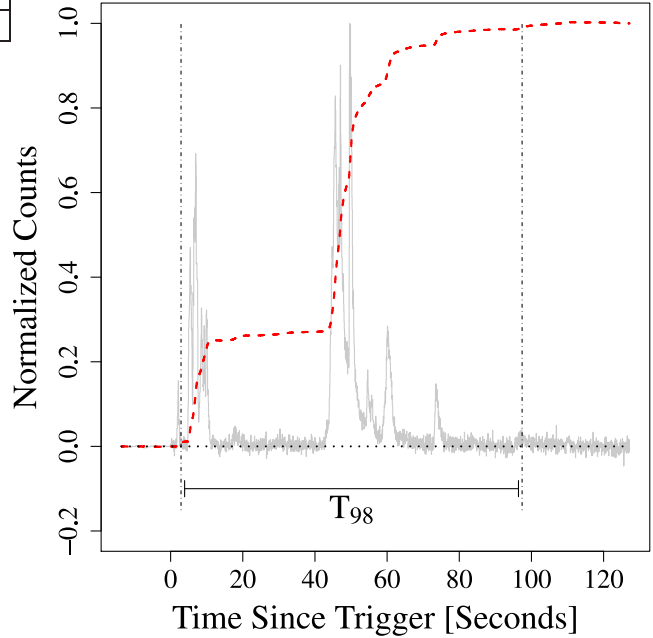}
\end{center}
\caption{An example of the cumulative light curve of GRB 990506.
The cumulative light curve is plotted with a red dashed line, and the
ordinary light curve with a gray solid line.
The vertical lines show the start and end time of $T_{\rm 98}$.
The interval within $T_{98}$ is indicated by an arrow.}
\label{fig:lightcurve}
\end{figure}

Because the observed distribution of GRBs spans the wide range of redshift,
from 0.0085 to 8.2 up to present, and the cosmological time dilation
changes the time
scales of bursts by factor $(1+z)$,
the measurement of temporal properties of bursts must be robust to the
change of time resolution of light curves.
Besides, recent results from hydrodynamical simulations of jet imply
that the long term variability of GRBs come from the activity of the
central engine, while the short term variability from the propagation of
jet \citep{Morsony:2010} and we should decompose long and short term
variability in some way.
For these reasons, we use cumulative light curves to quantify the
difference of bursts which are
much simpler than ordinary light curves \citep{McBreen:2002,Varga:2005}.
Figure \ref{fig:lightcurve} shows an example of the cumulative light
curve of GRB 990506.
The cumulative light curve normalized with total counts is plotted with
a red dashed line,
and the ordinary light curve normalized with peak counts with a gray
solid line, respectively.
The short scale structure in the ordinary light curve is canceled out in
the cumulative light curve, and then the analysis of cumulative light
curves is much robust to a cosmological time dilation and choice of the
time resolution.

\begin{figure}[htbp]
\begin{center}
\FigureFile(85mm,85mm){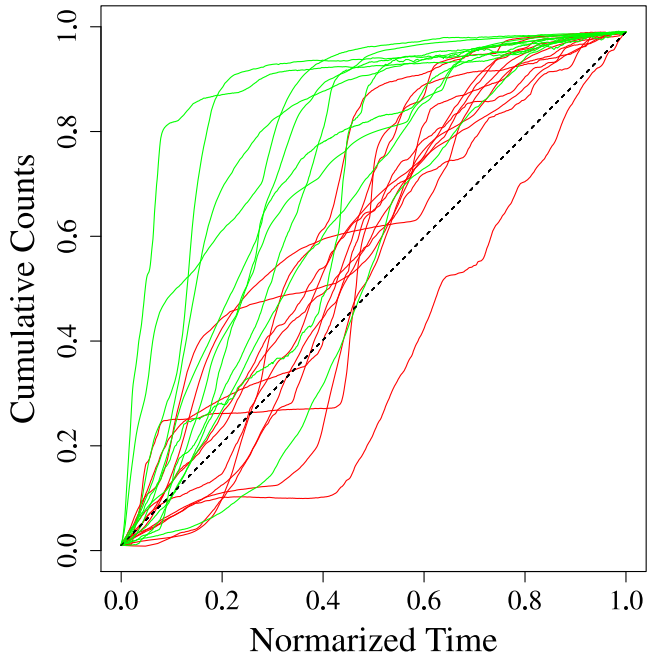}
\end{center}
\caption{The normalized cumulative light curves for the platinum samples.
On-plane events are plotted with red solid
lines,  outliers with green solid lines.
The black dotted line is the line of the virtual source with the
constant luminosity.
}
\label{fig2}
\end{figure}

To compare pulse shapes of bursts with various durations, we normalize
the time scale of bursts with $T_{98}$
and then all bursts have normalized cumulative counts $C_{i}^{\rm norm}$
from 0.01 to 0.99, and normalized time $t_{i}^{\rm norm}$ from 0 to 1.
Figure \ref{fig2} shows normalized cumulative light curves
of the platinum samples in table \ref{tab:intrinsic}.
On-plane events are plotted with red lines, outliers with green lines in
figure \ref{fig2}.
The black dotted line is the line of the virtual source with the
constant luminosity.
The brighter and dimmer outliers deviate from the constant luminosity line,
while many of on-plane events cluster around the constant luminosity line.

\begin{figure}
\begin{center}
\FigureFile(85mm,85mm){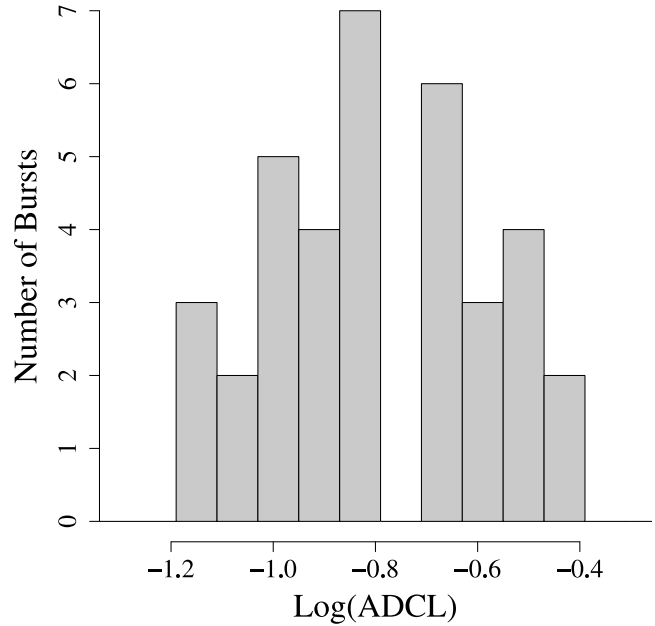}
\end{center}
\caption{The distribution of $ADCL$. There is a gap around $log
(ADCL)\sim -0.75$.}
\label{fig:ADCL}
\end{figure}

\begin{figure*}[hb]
\begin{center}
\begin{tabular}{cc}
\FigureFile(85mm,85mm){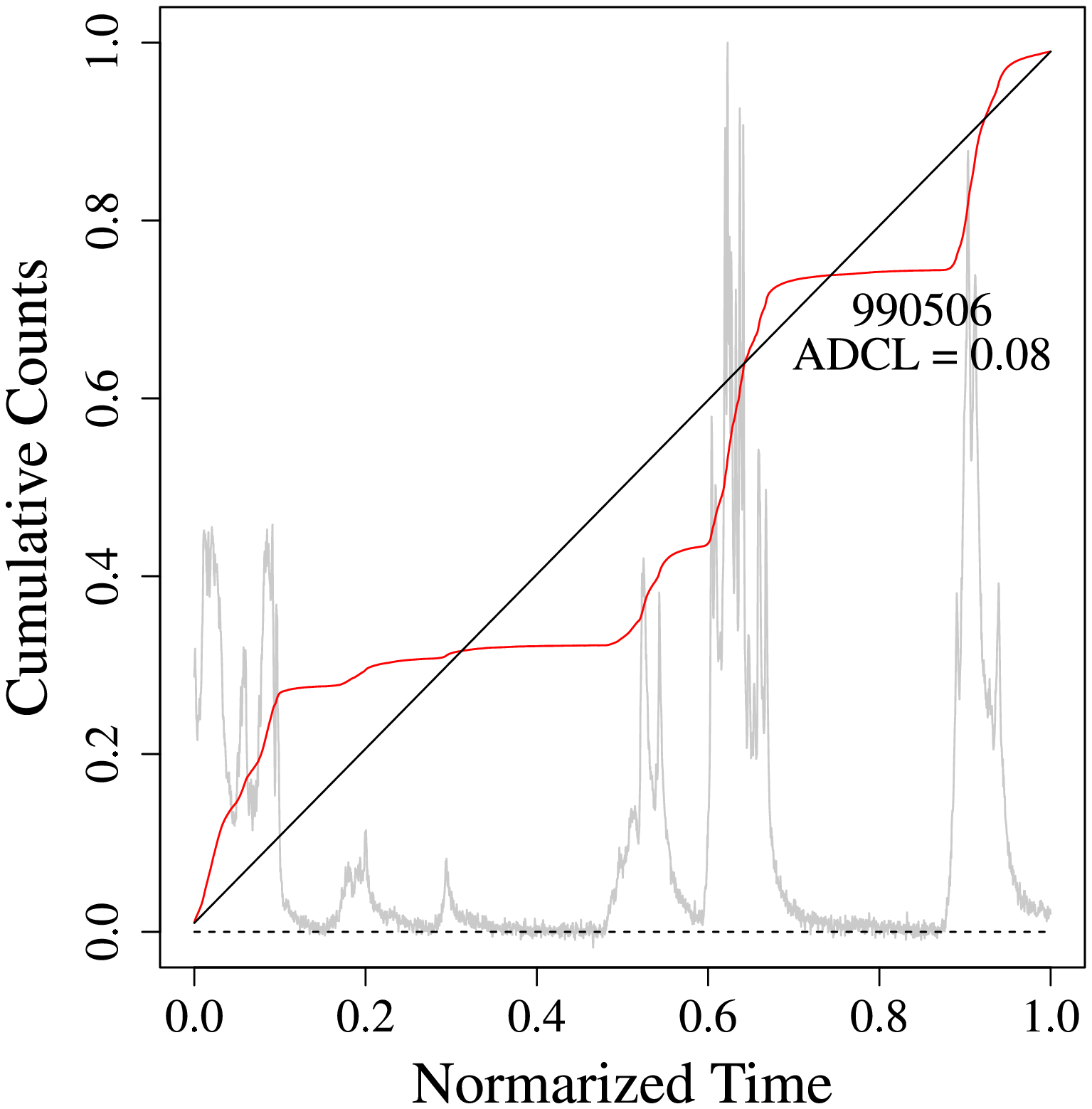}&
\FigureFile(85mm,85mm){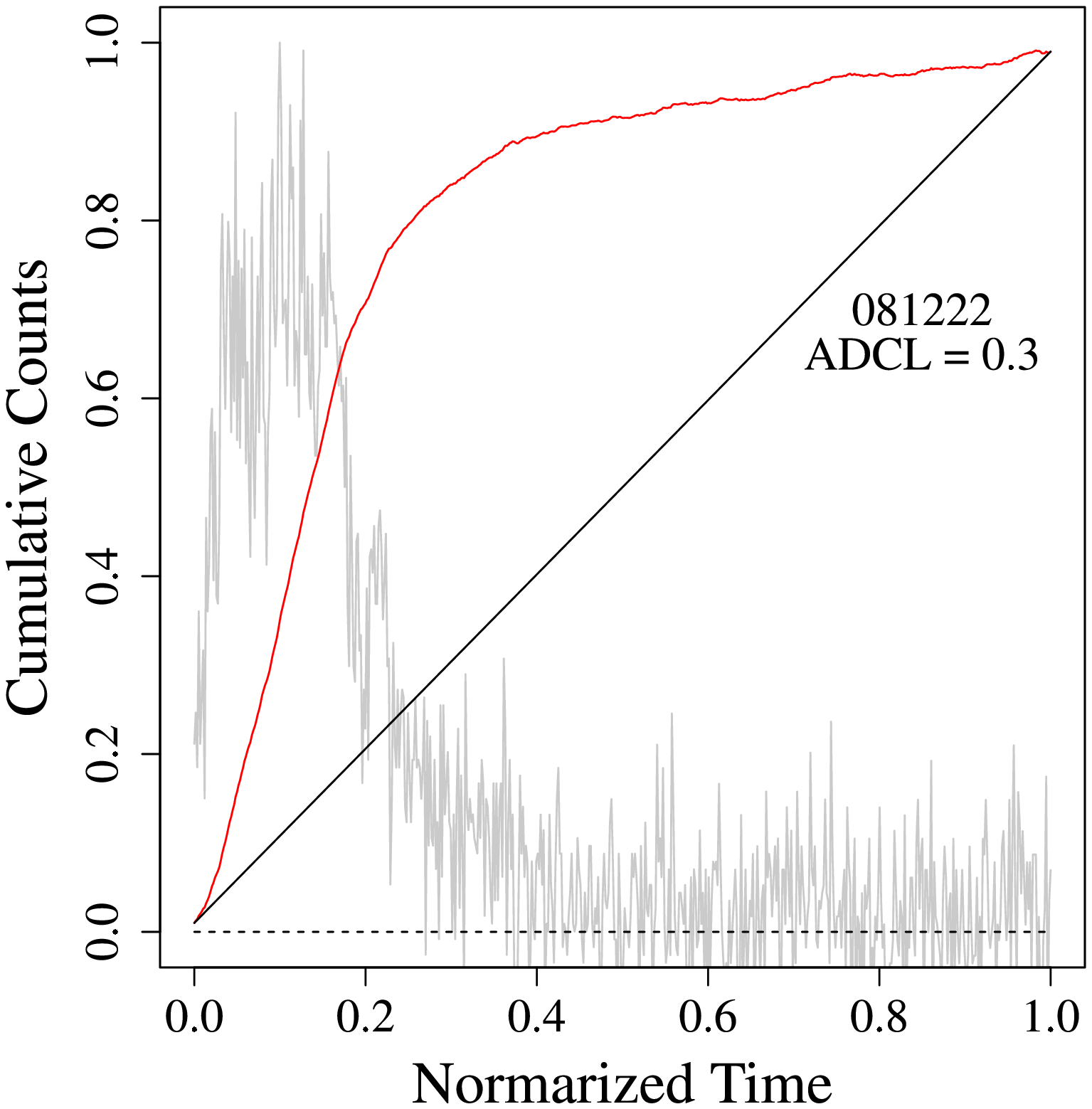}
\end{tabular}
\end{center}
\caption{{\it left} : An example of a cumulative light curve of
small-$ADCL$ GRB. {\it right} : An example of a cumulative light curve of
long tailed GRB.}
\label{fig:ADCL2}
\end{figure*}

To quantify these difference, we introduce a new parameter of GRBs :
the absolute deviation from their constant luminosity ($ADCL$).
We define the absolute deviation from their constant luminosity :
\begin{equation}
ADCL = \sum_{i=1}^{N_{\rm bin}} \frac{|C_{i}^{\rm norm}-0.01-0.98\times
t_i^{\rm
norm}|}{N_{\rm bin}}
\end{equation}
where the number of the bin ($N_{\rm bin}$) is different from burst to
burst. Figure \ref{fig:ADCL} shows the distribution of $ADCL$.
The distribution of $ADCL$ in figure \ref{fig:ADCL} indicates that there
is a gap around
$\log(ADCL)\sim-0.75~ (ADCL\sim0.17)$.
From this result, we divide the LGRBs into two subclasses : one is
small-$ADCL$ events with $ADCL<0.17$ and the other is long tailed ones with
$ADCL>0.17$.
Figure \ref{fig:ADCL2} shows examples of cumulative light curves of
small-$ADCL$ (left) and long tailed (right) events, respectively.
Although it is difficult to quantify the statistical significance of the
existence of the gap, we will show that the gap becomes more clear
when we study the relationship between the $ADCL$ and residual
from the Fundamental Plane of LGRBs. 
This is very similar to the discovery of the short - hard and long - soft
classification by \citet{Kouveliotou:1993}.
Only from $T_{\rm 90}$, the long and short classification is not so
clear  that
they added the spectral hardness. Then they could identify  the bimodal
distribution of the long-soft and short-hard GRBs clearer.
In our case, the residual from  the Ep-TL-Lp correlation corresponds to
the hardness.
In  Figure 8. we can identify two suclasses (blue squares and red
circles) in the ADCL-residual
plane.    

In table \ref{tab:intrinsic}, we summarize the intrinsic properties and
$ADCL$ of the 36
GRBs.


\subsection{Type I Fundamental plane}

In \citet{Tsutsui:2011} and \S-3, we have found that there are probably
three classes
of LGRBs : on-plane events, brighter outliers, and dimmer outliers.
In section \ref{sec:ADCL}, we introduced a new parameter $ADCL$ and found
that the most of on-plane events have smaller $ADCL$, while most of
the brighter and dimmer outliers have relatively larger $ADCL$.
In this section, we derive the best fit function and dispersion of
the \tsutsui correlation for small-$ADCL$ and large-$ADCL$ GRBs
separately, using the same method in section 3.

We used only the platinum data with
$\sigma_{E_{\rm p}}/E_{\rm p}<0.10$ in \citet{Tsutsui:2011} and section 3
to avoid the contamination of outliers and subclasses because there are
no prior
knowledge about outliers and subclasses of LGRBs. Now ADCL is  a
candidate for another
observational property to distinguish outliers and  subclass of LGRBs.
Using ADCL
we might eliminate brighter and dimmer outliers in advance.
Therefore we do not need to restrict the sample to the platinum one from
now on.
The platinum sample was needed only to find ADCL so that  we will use
the whole sample
in the following analysis although larger observational errors compared
with the platinum sample exist.

First, we apply the outlier elimination technique to the small-$ADCL$
GRBs with $ADCL < 0.17$ and find the \tsutsui correlation of
small-$ADCL$ GRBs becomes tightest around $\tau =2.752$ seconds.
The best fit relation is given by,
\begin{eqnarray}
\label{eq:FP1}
L_{\rm p}=10^{52.53\pm 0.01}&&
\left(\frac{E_{\rm p}}{10^{2.71}
{\rm keV}}\right)^{1.84\pm 0.03} \nonumber \\
&& \times \left(\frac{T_{\rm L}}{10^{0.86}{\rm sec}}\right)^{0.29\pm0.08},
\end{eqnarray}
with $\chi^2_{\nu}=10.93/14$ and $\sigma_{\rm int}=0$.
The power-law index of $T_L$ in equation.
(\ref{eq:FP1}) is different from that of the equation. (12) of Tsutsui
et al., (2011). This is because there were some contamination of long
tailed GRBs for the lack of population of LGRBs.
Hereafter we refer to the equation (\ref{eq:FP1}) as the Type I
Fundamental Plane.

\begin{figure}[htb]
\begin{center}
\FigureFile(85mm,85mm){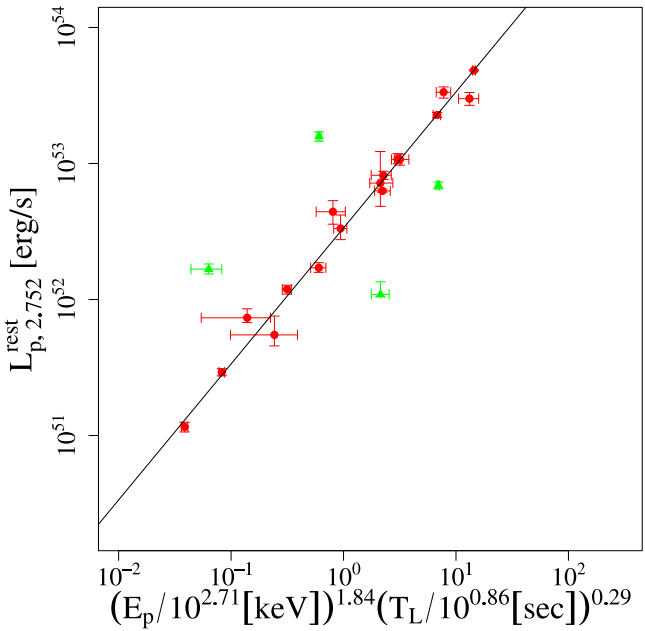}
\end{center}
\caption{The Type I Fundamental Plane of small-$ADCL$ events ($ADCL<0.17$).
The red circles represent the small-ADCL event used to derive the
Type I Fundamental Plane and the green triangles represent outliers
eliminated from chi square analysis.
The solid line represents the equation (\ref{eq:FP1}).
}
\label{fig:FP1}
\end{figure}

In figure \ref{fig:FP1}, we plot the data of the small-$ADCL$ events
with $ADCL<0.17$ and the best-fit function of the Type I Fundamental
Plane. The small-$ADCL$ events used to derive the best-fit function
were marked with red circles and the events regarded as outliers
with green triangles. The solid line represents the best-fit
function of the Type I Fundamental Plane.

\subsection{Type II Fundamental Plane}
We find that large-$ADCL$ GRBs form the other Fundamental Plane
which is almost parallel to Type I Fundamental Plane but slightly
brighter than it.
Furthermore, we find that all GRBs which have large $T_{\rm L} > 12$ sec
deviate from other events, and we eliminate them from following analysis.
Then we apply the outlier elimination technique to the long tailed
events with $T_{\rm L} < 12$ sec and estimate the best fit parameters
and resolution time scale of long tailed events.
We find the \tsutsui correlation of the long tailed events also becomes
tightest around 2.7 seconds. Adopting 2.752 seconds as the resolution
time scales of large-$ADCL$ events, we obtain
\begin{eqnarray}
\label{eq:FP2}
L_{\rm p}=10^{52.98\pm 0.080}&&
\left(\frac{E_{\rm p}}{10^{2.71}{\rm keV}}\right)^{1.82\pm
0.093}\nonumber \\
& &\times \left(\frac{T_{\rm L}}{10^{0.86}{\rm sec}}\right)^{0.85\pm0.26},
\end{eqnarray}
with $\chi^2_{\nu}=7.58/8$ and $\sigma_{\rm int}=0$ for long tailed GRBs.
Hereafter we refer to the equation (\ref{eq:FP2}) as the Type II Fundamental
Plane. The Type II Fundamental Plane is about 2.5 times brighter than
the Type I Fundamental Plane.

\begin{figure*}[hb]
\begin{center}
\begin{tabular}{cc}
\FigureFile(85mm,85mm){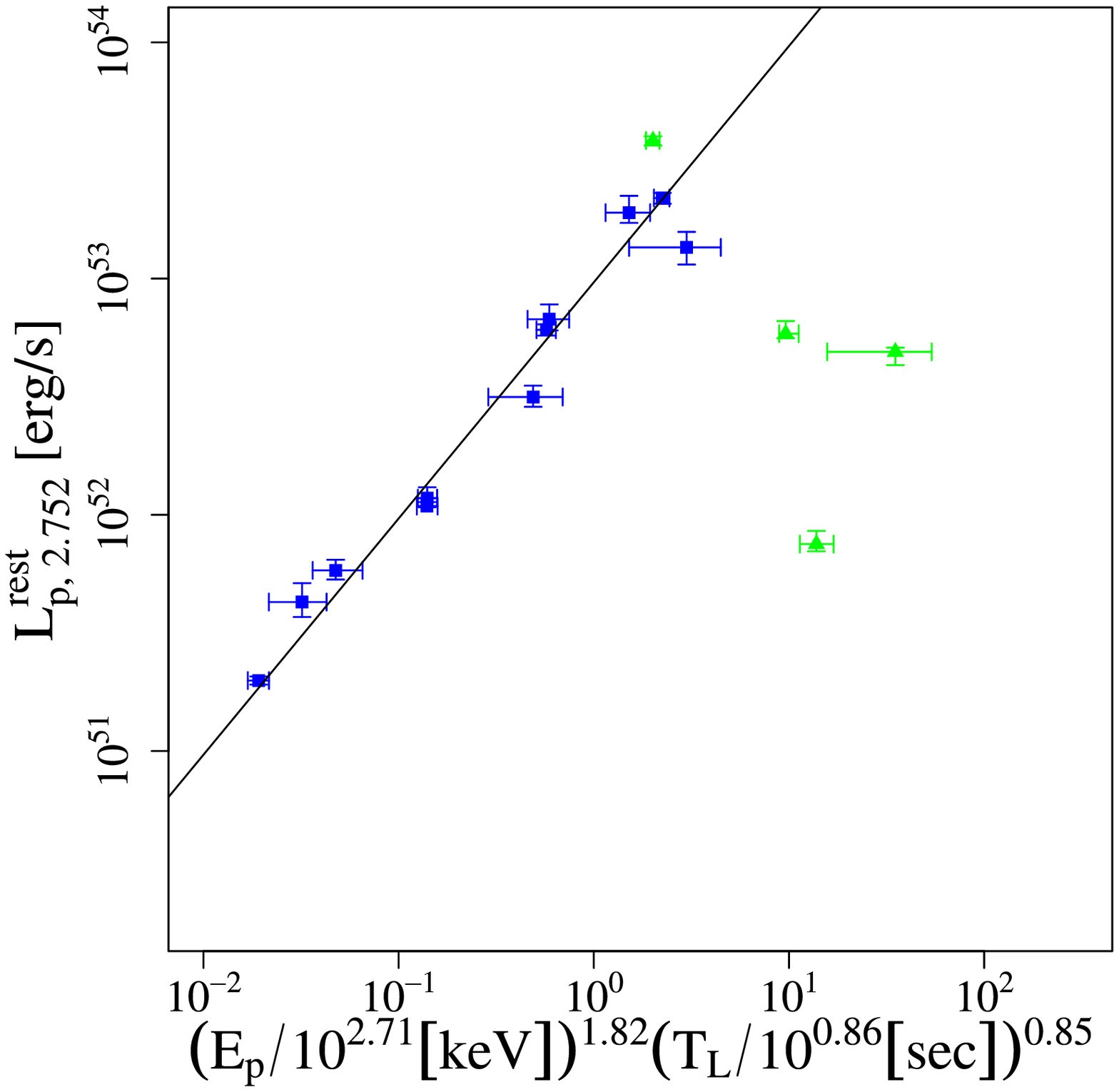}&
\FigureFile(85mm,85mm){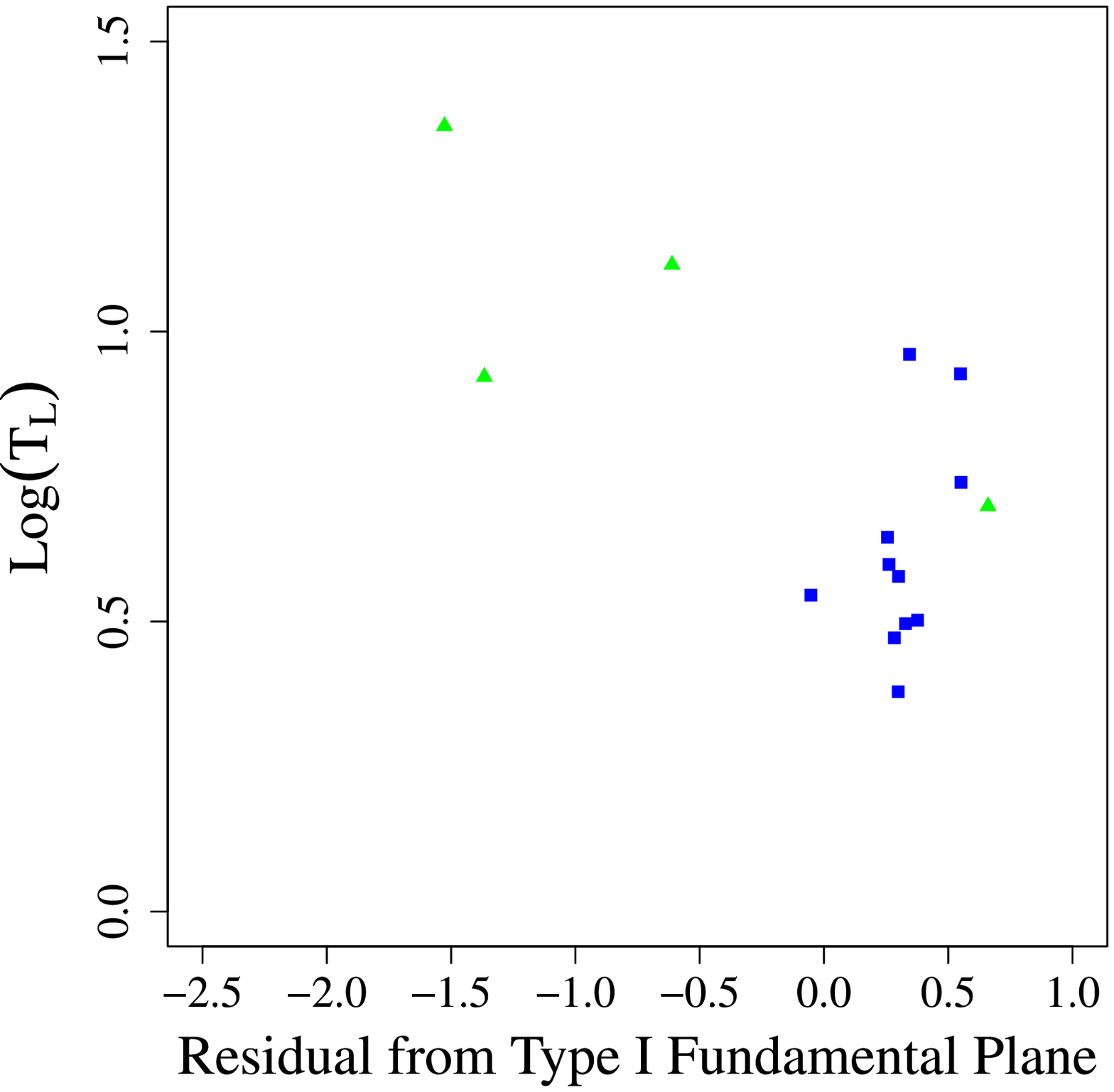}
\end{tabular}
\end{center}
\caption{{\it left} : The Fundamental Plane of large-$ADCL$ events.
The blue squares represent the events with $T_{\rm L} < 12$ sec used
to estimate the Type II Fundamental Plane and the green triangles
outliers eliminated from chi square analysis.
The solid line represents the equation (\ref{eq:FP2}).
{\it right} : The relationship between residual from the equation
(\ref{eq:FP2}) and $T_{\rm L}$ of the long tailed events.
Each symbols represents same as left figure.
}
\label{fig:FP2}
\end{figure*}

In figure \ref{fig:FP2}, we plot the data of large-$ADCL$ events
and the best-fit function of the Type II Fundamental Plane.
The events used to derive the best-fit function were marked with blue
squares and the events regarded as outliers with green triangles.
The solid line represents the best-fit function of the Type II Fundamental
Plane.

\begin{figure}[ht]
\begin{center}
\FigureFile(85mm,85mm){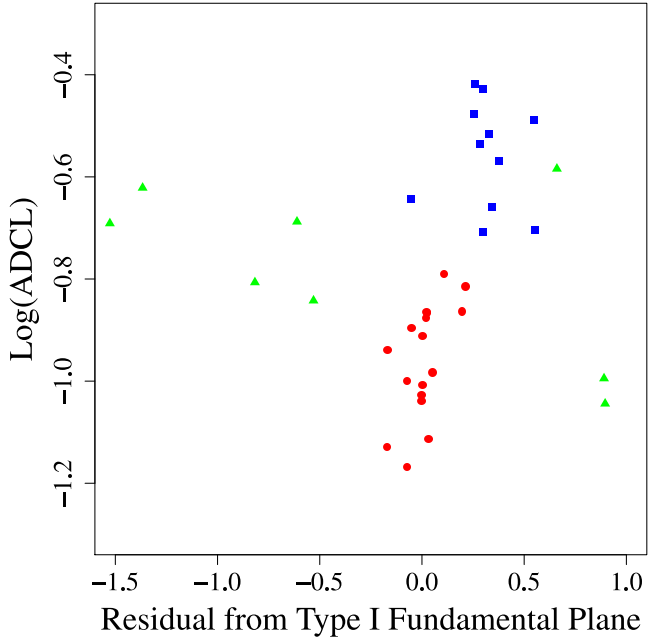}
\end{center}
\caption{The relationship between the residual from Type I fundamental plane
and $ADCL$ of all events. Each symbol represents same as figure
\ref{fig:FP1} and \ref{fig:FP2}.}
\label{fig:classification}
\end{figure}

Figure \ref{fig:classification} shows the relationship between residual
from the Type I fundamental plane  and $ADCL$ of all events. Each symbol
represents same as figure \ref{fig:FP1} and \ref{fig:FP2}.
Dimmer outliers apparently have intermediate $ADCL$ value.
Figure \ref{fig:classification} might imply that there are some other
classes of LGRBs.

In section 3 we showed that the \tsutsui correlation for whole sample
has significant intrinsic dispersion
($\sigma_{\rm int}=0.13$ for whole sample while 0.05 for platinum
sample. see Table. 1).
If we  divide the whole sample into two subclasses from their $ADCL$ values,
$\sigma_{\rm int}=0$ as shown in this chapter.
Thus we can conclude that the \tsutsui correlation is certainly improved
by introducing $ADCL$.

\section{Discussion}
In this section, we would like to show the validity of the \tsutsui
correlation.
First one might ask that  $T_{\rm L}$ might introduce spurious correlation
because $T_{\rm L}$ includes $L_{\rm p}$ in its definition. We checked this
by assuming the relation as
$\log L_{\rm p}(E_{\rm p},E_{\rm iso})= {A^{'}} +B^{'}\log ({E_{\rm
p}}/{10^{2.71}{\rm keV}})+ C^{'} \log ({E_{\rm iso}}/{10^{53.40}{\rm
sec}})$.
We determined $A^{'}$, $B^{'}$ and $C^{'}$using the same
data and the same method. If there is no spurious correlation,
we should obtain $B^{'}=B/(1+C)$ and $C^{'}=C/(1+C)$ where A, B and C
are defined in Eq. (4).
We obtained that $B^{'}=1.42$, $C^{'}=0.23$ with $\chi_{\nu}^2=10.93/14$.
$B=1.84$, $C=0.29$ so that $B/(1+C)=1.42$, $C/(1+C)=0.22$ for
small-$ADCL$ sample
and that $B^{'}=0.98$, $C^{'}=0.46$ with $\chi_{\nu}^2=7.59/9$.
$B=1.82$, $C=0.85$ so that $B/(1+C)=0.98$, $C/(1+C)=0.46$ for large ADCL
sample.
These values show that the spurious correlation is very small even if it
exists. Therefore which one should be used is a matter of taste. 

Next we compare  the \tsutsui correlation with \yonetoku  correlation
for each subsample with the same statistical method in section 3.
We first  derive the \yonetoku correlation with the whole sample as
discussed in section 3.
In table \ref{tab:Ep-Lp} we summarize  the best fit parameters, fraction
of outliers ($N_{\rm out}/N$),
intrinsic dispersion ($\sigma_{\rm int}$) and reduced chi square
($\chi_{\nu}^2$)
both for the whole sample and platinum sample.
As table \ref{tab:allsample} and \ref{tab:Ep-Lp} indicate the \yonetoku
correlation
has the larger systematic dispersion  than the \tsutsui correlation when
we use the whole sample.
the whole sample.

Secondly  we divide the GRBs into two groups by $ADCL$, and derive the
\yonetoku correlation for each sample.
Here we should note without the \tsutsui correlation it was not possible
to find ADCL and to recognize the existence of
the subclasses. However now we know that two subclasses exist so that we
check the \yonetoku correlation for each subclass.
In table \ref{tab:small-ADCL} and \ref{tab:longtailed}, we summarize
the best fit parameters, fraction of outliers ($N_{\rm out}/N$),
intrinsic dispersion ($\sigma_{\rm int}$)
and reduced chi square ($\chi_{\nu}^2$) for each correlation and each
subclass.
As table  \ref{tab:small-ADCL} and \ref{tab:longtailed} indicate, the
difference in parameters  of the \yonetoku correlation
between small-$ADCL$ events and large-$ADCL$ (long tailed) events is
smaller than that of the \tsutsui correlation. Therefore
we can not distinguish them even if we use our outlier rejection
technique since the difference of the amplitude A is almost the same ($\sim
0.2 $  in logarithm \yonetoku correlation while $\sim 0.45$ for
\tsutsui correlation) . Therefore \tsutsui correlation is needed
to distinguish two subclasses.
In case of small-$ADCL$ events, the \yonetoku correlation has slightly
larger dispersion than the \tsutsui correlation,
although the \yonetoku correlation has one more outlier.
Taking into these facts, we can insist that the
\tsutsui correlation is surely more statistically significant than
\yonetoku correlation
for small-$ADCL$ events.
In case of long tailed events, however, it is difficult to say which one
is more significant.
The \yonetoku correlation has two more outliers, but smaller dispersion
than the \tsutsui correlation.
To show the validity of the \tsutsui correlation for long tailed events,
we need larger number of data.

\begin{table}[htdp]
\caption{Fitting results for the \yonetoku correlation with the whole
sample.}
\begin{center}
\begin{tabular}{|c|c|c|c|c|}
\hline
sample &  best fit  & $\sigma_{\rm int}$ & $N_{\rm out}$/$N$ &
$\chi^2_{\nu}$ \\ \hline
all & $(52.64, 1.62)$  & 0.18 & 6/36 & 27.6/28 \\ \hline
platinum & $(52.57, 1.71)$ 	&  0.15	& 6/23 & 12.9/15 \\ \hline
\end{tabular}
\end{center}
\label{tab:Ep-Lp}
\end{table}

\begin{table*}[htdp]
\caption{Fitting results for small-$ADCL$ events.}
\begin{center}
\begin{tabular}{|c|c|c|c|c|c|}
\hline
correlation & best fit  & $N_{\rm out}/N$ &$\sigma_{\rm int}$ &
$\chi^2_{\nu}$ \\ \hline
\yonetoku & $(52.54, 1.81)$ & 5/21 & 0 & 15.55/14 \\ \hline
\tsutsui & $(52.53,1,84,0.29)$ & 4/21 & 0 &  10.93/14\\ \hline
\end{tabular}
\end{center}
\label{tab:small-ADCL}
\end{table*}%

\begin{table*}[htdp]
\caption{Fitting results for long tailed events.}
\begin{center}
\begin{tabular}{|c|c|c|c|c|}
\hline
correlation & best fit  & $N_{\rm out}$/N &$\sigma_{\rm int}$
&$\chi^2_{\nu}$  \\ \hline
\yonetoku & $(52.73, 1.78)$& 6/15 & 0 & 4.672/7 \\ \hline
\tsutsui & $(52.98,1,82,0.85)$ & 4/15 & 0 & 7.58/8 \\ \hline
\end{tabular}
\end{center}
\label{tab:longtailed}
\end{table*}%


\section{Summary}
In this paper we defined $ADCL$ which characterizes the cumulative light
curve
of the prompt emission of long GRBs. We divided the events into two groups
according to the value of $ADCL$. Then we derived Fundamental Planes for
the two groups separately and it was found that they are tighter and
statistically more significant than one derived from the whole sample.
Although we introduced the new parameter $ADCL$ to
reduce the dispersion of the Fundamental Plane, we found dividing the
events into two groups
also improves the \yonetoku correlation. This fact supports the validity
of introducing $ADCL$.
 These tighter relations would be helpful to use them as standard candles
to probe cosmological expansion history. Also our analysis may imply
the existence of subclasses of long GRBs classified by $ADCL$.

The discovery of these possible subclasses of GRBs reminds us of the
discovery of the two separate subclasses of Cepheids by Baade.
After Baade's discovery, the accuracy of the distance measurement by
the Period-Luminosity relation of Cepheids was drastically improved.
We can expect that the distance measurement by
the $E_{\rm p}$--$T_{\rm L}$--$L_{\rm p}$ relation of GRBs will also
be improved, and we are studying the distance measurement by the Type I
Fundamental Plane of GRBs in near future.

\section*{Acknowledgments}

This work is supported in part by the Grant-in-Aid from the
Ministry of Education, Culture, Sports, Science and Technology
(MEXT) of Japan, No.23540305 (TN), No.20674002 (DY), No.23740179 (KT),
and by the Grant-in-Aid for the global COE program
{\it The Next Generation of Physics, Spun from Universality
and Emergence} at Kyoto University.


\begin{table*}[htbp]
\begin{minipage}[t]{1 \hsize}
{\scriptsize
{\tabcolsep = 0.5mm
\caption{Spectropic redshifts, spectral parameters determined by the Band model, observed fluxes and fluences with energy band of detectors, and references of 44 LGRBs }
\begin{tabular}{l|cccccccc}
\hline
GRB	&	$z$\footnote{Taken from the J. Greiner's GRB table and references therein (http://www.mpe.mpg.de/~jcg/grbgen.html).}	&	$\alpha$	&$E_{\rm p}^{obs}$	&$\beta$ 
& $P_{\rm p}^{\rm obs}$($E_1-E_2$)\footnote{Observed peak photon (or energy) fluxes  between $E_1$ to $E_2$ keV.}&	$S_{\rm obs} $	($E_3-E_4$)\footnote{Observed fluences  between $E_3$ to $E_4$ keV.}
&detector\footnote{Detectors that provide spectral parameters and light curves (SAX={\it BeppoSAX}, K={\it Konus}, BAT=BATSE, HE= HETE-II, S={\it Swift}, SK = {\it Swift} and {\it Konus}, SW= {\it Swift} and WAM)} &	
 reference\footnote{ References for spectral parameters, peak fluxes and fluences : (1) \citet{Amati:2002} ; (2) \citet{Yonetoku:2004} ; (3) \citet{Sakamoto:2005} ; (4) \citet{Krimm:2006a} ; (5) \citet{GCN5722} ; (6) \citet{GCN6049} ;	(7) \citet{Krimm:2009} ; (8) \citet{GCN7482} ; (9) \citet{GCN7995} ; (10) This work ; (11) \citet{GCN8548} ; 
 (12) \citet{GCN9030} ; (13) \citet{GCN11971} ; (14) \citet{GCN12008} ; (15)  \citet{GCN12166}
}\\
	&		&		&
	     [keV]	&
		 &					[(erg or
		     photon)/cm$^{-2}$s$^{-1}$]	&
			 [erg/cm$^{-2}$]	& 
			     &		\\ \hline
970228	&	0.695	& $	-1.54	^{+	0.08	}_{	-0.08	}$& $	115.00	^{+	38.00	}_{	-38.00	}$&$	-2.50	^{+	0.40	}_{	-0.40	}$&$							(3.70	^{+	0.10	}_{	-0.10	})\times 10^{	-6	}$(	40	-	700	)&$(	1.10	^{+	0.10	}_{	-0.10	})\times 10^{	-5	}$(	40	-	700	)&	SAX/K	&	1	\\
971214	&	3.418	& $	-0.36	^{+	0.14	}_{	-0.14	}$& $	182.60	^{+	11.00	}_{	-14.30	}$&$	-2.10	^{+	0.52	}_{	-0.90	}$&$	1.95	^{+	0.05	}_{	-0.05	}$(									50	-	300	)&$(	4.96	^{+	0.07	}_{	-0.07	})\times 10^{	-6	}$(	100	-	300	)&	BAT	&	2	\\
980425	&	0.0085	& $	-0.97	^{+	0.16	}_{	-0.16	}$& $	54.90	^{+	11.50	}_{	-11.50	}$&$	-2.06	^{+	0.09	}_{	-0.09	}$&$	0.96	^{+	0.05	}_{	-0.05	}$(									50	-	300	)&$(	1.67	^{+	0.07	}_{	-0.07	})\times 10^{	-6	}$(	100	-	300	)&	BAT	&	2	\\
990123	&	1.6	& $	-0.18	^{+	0.08	}_{	-0.07	}$& $	513.00	^{+	19.20	}_{	-21.90	}$&$	-2.33	^{+	0.08	}_{	-0.09	}$&$	16.41	^{+	0.12	}_{	-0.12	}$(									50	-	300	)&$(	8.72	^{+	0.02	}_{	-0.02	})\times 10^{	-5	}$(	100	-	300	)&	BAT	&	2	\\
990506	&	1.3	& $	-0.9	^{+	0.19	}_{	-0.13	}$& $	320.70	^{+	30.10	}_{	-38.20	}$&$	-2.08	^{+	0.08	}_{	-0.10	}$&$	18.56	^{+	0.13	}_{	-0.13	}$(									50	-	300	)&$(	5.16	^{+	0.02	}_{	-0.02	})\times 10^{	-5	}$(	100	-	300	)&	BAT	&	2	\\
990510	&	1.619	& $	-0.71	^{+	0.12	}_{	-0.12	}$& $	205.50	^{+	9.60	}_{	-12.30	}$&$	-2.79	^{+	0.51	}_{	-6.21	}$&$	8.17	^{+	0.08	}_{	-0.08	}$(									50	-	300	)&$(	8.04	^{+	0.08	}_{	-0.08	})\times 10^{	-6	}$(	100	-	300	)&	BAT	&	2	\\
990705	&	0.843	& $	-1.05	^{+	0.21	}_{	-0.21	}$& $	189.00	^{+	15.00	}_{	-15.00	}$&$	-2.20	^{+	0.10	}_{	-0.10	}$&$							(3.70	^{+	0.10	}_{	-0.10	})\times 10^{	-6	}$(	40	-	700	)&$(	7.50	^{+	0.80	}_{	-0.80	})\times 10^{	-5	}$(	40	-	700	)&	SAX/K	&	1	\\
990712	&	0.434	& $	-1.88	^{+	0.07	}_{	-0.07	}$& $	65.00	^{+	11.00	}_{	-11.00	}$&$	-2.48	^{+	0.56	}_{	-0.56	}$&$							(1.30	^{+	0.10	}_{	-0.10	})\times 10^{	-6	}$(	40	-	700	)&$(	6.50	^{+	0.30	}_{	-0.30	})\times 10^{	-6	}$(	40	-	700	)&	SAX/K	&	1	\\
991216	&	1.02	& $	-0.66	^{+	0.04	}_{	-0.04	}$& $	536.50	^{+	18.50	}_{	-20.40	}$&$	-2.44	^{+	0.12	}_{	-0.17	}$&$	67.52	^{+	0.23	}_{	-0.23	}$(									50	-	300	)&$(	6.37	^{+	0.01	}_{	-0.01	})\times 10^{	-5	}$(	100	-	300	)&	BAT	&	2	\\
021211	&	1.01	& $	-0.86	^{+	0.10	}_{	-0.09	}$& $	45.56	^{+	7.84	}_{	-6.23	}$&$	-2.18	^{+	0.14	}_{	-0.25	}$&$	30.00	^{+	1.32	}_{	-1.32	}$(									2	-	400	)&$(	3.53	^{+	0.21	}_{	-0.21	})\times 10^{	-6	}$(	2	-	400	)&	HE/K	&	3	\\
030329	&	0.168	& $	-1.26	^{+	0.01	}_{	-0.02	}$& $	67.86	^{+	2.31	}_{	-2.15	}$&$	-2.28	^{+	0.05	}_{	-0.06	}$&$	72.21	^{+	2.86	}_{	-2.86	}$(									30	-	400	)&$(	1.08	^{+	0.01	}_{	-0.01	})\times 10^{	-4	}$(	30	-	400	)&	HE/K	&	3	\\
050401	&	2.9	& $	-0.90	^{+	0.30	}_{	-0.30	}$& $	117.50	^{+	18.00	}_{	-18.00	}$&$	-2.55	^{+	0.22	}_{	-0.44	}$&$	10.70	^{+	0.92	}_{	-0.92	}$(									15	-	150	)&$(	8.22	^{+	0.16	}_{	-0.16	})\times 10^{	-6	}$(	15	-	150	)&	SK	&	4	\\
050525	&	0.606	& $	-1.01	^{+	0.11	}_{	-0.11	}$& $	81.20	^{+	2.30	}_{	-2.30	}$&$	-3.26	^{+	0.23	}_{	-0.41	}$&$	41.70	^{+	0.94	}_{	-0.94	}$(									15	-	150	)&$(	1.53	^{+	0.02	}_{	-0.02	})\times 10^{	-5	}$(	15	-	150	)&	SK	&	4	\\
050603	&	2.821	& $	-1.03	^{+	0.11	}_{	-0.11	}$& $	343.7	^{+	87.00	}_{	-87.00	}$&$	-2.03	^{+	0.17	}_{	-0.29	}$&$	21.50	^{+	1.07	}_{	-1.07	}$(									15	-	150	)&$(	6.36	^{+	0.23	}_{	-0.23	})\times 10^{	-6	}$(	15	-	150	)&	SK	&	4	\\
061007	&	1.261	& $	-0.7	^{+	0.02	}_{	-0.02	}$& $	399	^{+	10.94	}_{	-11.55	}$&$	-2.61	^{+	0.09	}_{	-0.13	}$&$							(1.66	^{+	0.16	}_{	-0.12	})\times 10^{	-5	}$(	20	-	10000	)&$(	2.49	^{+	0.10	}_{	-0.07	})\times 10^{	-4	}$(	20	-	10000	)&	K	&	5	\\
070125	&	1.547	& $	-1.1	^{+	0.06	}_{	-0.05	}$& $	367.00	^{+	39.51	}_{	-31.00	}$&$	-2.08	^{+	0.06	}_{	-0.09	}$&$							(1.41	^{+	0.14	}_{	-0.14	})\times 10^{	-5	}$(	20	-	10000	)&$(	1.74	^{+	0.11	}_{	-0.09	})\times 10^{	-4	}$(	20	-	10000	)&	K	&	6	\\
071003	&	1.6044	& $	-1.22	^{+	0.04	}_{	-0.04	}$& $	1,307.00	^{+	381.00	}_{	-381.00	}$&$	-9.27	^{+	7.14	}_{	-0.73	}$&$	6.30	^{+	0.24	}_{	-0.24	}$(									15	-	150	)&$(	8.30	^{+	0.18	}_{	-0.18	})\times 10^{	-6	}$(	15	-	150	)&	SW/K	&	7	\\
071010B	&	0.947	& $	-1.34	^{+	0.47	}_{	-0.47	}$& $	45.00	^{+	4.00	}_{	-7.00	}$&$	-2.34	^{+	0.16	}_{	-0.26	}$&$	7.70	^{+	0.06	}_{	-0.06	}$(									15	-	150	)&$(	4.40	^{+	0.06	}_{	-0.06	})\times 10^{	-6	}$(	15	-	150	)&	SW/S	&	7	\\
080319B	&	0.937	& $	-0.822	^{+	0.01	}_{	-0.01	}$& $	651.00	^{+	7.90	}_{	-8.51	}$&$	-3.87	^{+	0.27	}_{	-0.66	}$&$							(1.67	^{+	0.10	}_{	-0.10	})\times 10^{	-5	}$(	20	-	7000	)&$(	5.72	^{+	0.09	}_{	-0.08	})\times 10^{	-4	}$(	20	-	7000	)&	K	&	8	\\
080413B	&	1.1	& $	-1.24	^{+	0.26	}_{	-0.26	}$& $	67.00	^{+	13.00	}_{	-8.00	}$&$	-2.77	^{+	0.22	}_{	-0.27	}$&$	18.70	^{+	0.49	}_{	-0.49	}$(									15	-	150	)&$(	3.20	^{+	0.06	}_{	-0.06	})\times 10^{	-6	}$(	15	-	150	)&	SW/S	&	7	\\
080721	&	2.602	& $	-0.933	^{+	0.06	}_{	-0.05	}$& $	485.00	^{+	40.73	}_{	-35.87	}$&$	-2.43	^{+	0.15	}_{	-0.26	}$&$							(1.15	^{+	0.11	}_{	-0.11	})\times 10^{	-5	}$(	20	-	5000	)&$(	8.38	^{+	0.38	}_{	-0.36	})\times 10^{	-5	}$(	20	-	5000	)&	K	&	9 	\\
080916A	&	0.689	& $	-1.11	^{+	0.11	}_{	-0.11	}$& $	129.30	^{+	23.30	}_{	-23.30	}$&$	-2.49	^{+	0.53	}_{	-0.53	}$&$	4.50	^{+	0.70	}_{	-0.70	}$(									25	-	1000	)&$(	1.50	^{+	0.50	}_{	-0.50	})\times 10^{	-5	}$(	25	-	1000	)&	GBM	&	10	\\
081121	&	2.512	& $	-0.77	^{+	0.09	}_{	-0.09	}$& $	248.00	^{+	23.10	}_{	-19.45	}$&$	-2.51	^{+	0.19	}_{	-0.40	}$&$							(1.94	^{+	0.03	}_{	-0.03	})\times 10^{	-6	}$(	20	-	7000	)&$(	1.79	^{+	0.22	}_{	-0.19	})\times 10^{	-5	}$(	20	-	7000	)&	K	&	11	\\
081222	&	2.77	& $	-0.91	^{+	0.07	}_{	-0.07	}$& $	150.50	^{+	15.80	}_{	-15.80	}$&$	-2.28	^{+	0.19	}_{	-0.19	}$&$	7.70	^{+	0.12	}_{	-0.12	}$(									15	-	150	)&$(	4.80	^{+	0.06	}_{	-0.06	})\times 10^{	-6	}$(	15	-	150	)&	GBM	&	10	\\
090323	&	3.57	& $	-0.96	^{+	0.07	}_{	-0.05	}$& $	416.00	^{+	46.20	}_{	-44.38	}$&$	-2.09	^{+	0.10	}_{	-0.13	}$&$							(5.17	^{+	0.57	}_{	-0.56	})\times 10^{	-6	}$(	20	-	10000	)&$(	2.02	^{+	0.17	}_{	-0.15	})\times 10^{	-4	}$(	20	-	10000	)&	K	&	12  	\\
090328	&	0.736	& $	-1.12	^{+	0.02	}_{	-0.02	}$& $	738.10	^{+	67.00	}_{	-67.00	}$&$	-2.77	^{+	0.49	}_{	-0.49	}$&$	18.50	^{+	0.50	}_{	-0.50	}$(									8	-	1000	)&$(	8.09	^{+	0.06	}_{	0.06	})\times 10^{	-5	}$(	8	-	1000	)&	GBM	&	10	\\
090424	&	0.544	& $	-0.9	^{+	0.03	}_{	-0.03	}$& $	149.00	^{+	5.08	}_{	-5.08	}$&$	-2.62	^{+	0.15	}_{	-0.15	}$&$	118.45	^{+	4.32	}_{	-4.32	}$(									8	-	1000	)&$(	5.20	^{+	0.10	}_{	-0.10	})\times 10^{	-5	}$(	8	-	1000	)&	GBM	&	10	\\
090618	&	0.54	& $	-0.87	^{+	0.11	}_{	-0.11	}$& $	149.10	^{+	7.16	}_{	-7.16	}$&$	-2.37	^{+	0.05	}_{	-0.05	}$&$	73.40	^{+	2.00	}_{	-2.00	}$(									8	-	1000	)&$(	2.70	^{+	0.06	}_{	-0.06	})\times 10^{	-4	}$(	8	-	1000	)&	GBM	&	10	\\
090902B	&	1.822	& $	-0.85	^{+	0.02	}_{	-0.02	}$& $	792.10	^{+	13.70	}_{	-13.70	}$&$	-3.80	^{+	0.25	}_{	-0.25	}$&$	46.10	^{+	0.30	}_{	-0.30	}$(									50	-	10000	)&$(	3.74	^{+	0.03	}_{	-0.03	})\times 10^{	-4	}$(	50	-	10000	)&	GBM	&	10	\\
090926B	&	1.24	& $	-0.11	^{+	0.16	}_{	-0.16	}$& $	88.55	^{+	5.41	}_{	-5.41	}$&$	-2.98	^{+	0.36	}_{	-0.36	}$&$	3.20	^{+	0.18	}_{	-0.18	}$(									15	-	150	)&$(	7.30	^{+	0.12	}_{	-0.12	})\times 10^{	-6	}$(	15	-	150	)&	GBM	&	10	\\
090926A	&	2.1062	& $	-0.95	^{+	0.06	}_{	-0.06	}$& $	288.30	^{+	9.78	}_{	-9.78	}$&$	-2.40	^{+	0.05	}_{	-0.05	}$&$	80.80	^{+	0.40	}_{	-0.40	}$(									8	-	1000	)&$(	1.45	^{+	0.04	}_{	-0.04	})\times 10^{	-4	}$(	8	-	1000	)&	GBM	&	10	\\
091003	&	0.8969	& $	-1.04	^{+	0.05	}_{	-0.05	}$& $	398.40	^{+	36.90	}_{	-36.90	}$&$	-3.58	^{+	1.21	}_{	-1.21	}$&$	31.80	^{+	0.40	}_{	-0.40	}$(									8	-	1000	)&$(	3.76	^{+	0.04	}_{	-0.04	})\times 10^{	-5	}$(	8	-	1000	)&	GBM	&	10	\\
091127	&	0.49	& $	-1.41	^{+	0.21	}_{	-0.21	}$& $	36.30	^{+	2.29	}_{	-2.29	}$&$	-2.20	^{+	0.02	}_{	-0.02	}$&$	46.90	^{+	0.90	}_{	-0.90	}$(									8	-	1000	)&$(	1.87	^{+	0.02	}_{	-0.02	})\times 10^{	-5	}$(	8	-	1000	)&	GBM	&	10	\\
091208B	&	1.063	& $	-1.44	^{+	0.15	}_{	-0.15	}$& $	124.70	^{+	40.00	}_{	-40.00	}$&$	-2.15	^{+	0.28	}_{	-0.28	}$&$	21.83	^{+	0.74	}_{	-0.74	}$(									8	-	1000	)&$(	5.80	^{+	0.20	}_{	-0.20	})\times 10^{	-6	}$(	8	-	1000	)&	GBM	&	10	\\
100414A	&	1.368	& $	-0.37	^{+	0.03	}_{	-0.03	}$& $	572.80	^{+	16.20	}_{	-16.20	}$&$	-3.74	^{+	0.95	}_{	-0.95	}$&$	18.22	^{+	0.24	}_{	-0.24	}$(									8	-	1000	)&$(	1.29	^{+	0.02	}_{	-0.02	})\times 10^{	-4	}$(	8	-	1000	)&	GBM	&	10	\\
100728B	&	2.106	& $	-0.93	^{+	0.24	}_{	-0.24	}$& $	115.30	^{+	29.20	}_{	-29.20	}$&$	-2.23	^{+	0.41	}_{	-0.41	}$&$	6.20	^{+	0.20	}_{	-0.20	}$(									8	-	1000	)&$(	2.40	^{+	0.10	}_{	-0.10	})\times 10^{	-6	}$(	8	-	1000	)&	GBM	&	10	\\
100814A	&	1.44	& $	-0.74	^{+	0.13	}_{	-0.13	}$& $	130.40	^{+	17.20	}_{	-17.20	}$&$	-2.73	^{+	0.69	}_{	-0.69	}$&$	4.50	^{+	0.20	}_{	-0.20	}$(									10	-	1000	)&$(	1.98	^{+	0.06	}_{	0.06	})\times 10^{	-5	}$(	10	-	1000	)&	GBM	&	10	\\
100906A	&	1.727	& $	-1.41	^{+	0.08	}_{	-0.08	}$& $	113.80	^{+	25.20	}_{	-25.20	}$&$	-2.01	^{+	0.09	}_{	-0.09	}$&$	14.45	^{+	0.29	}_{	-0.29	}$(									10	-	1000	)&$(	2.64	^{+	0.03	}_{	-0.03	})\times 10^{	-5	}$(	10	-	1000	)&	GBM	&	10	\\
101219B	&	0.55	& $	-0.63	^{+	0.73	}_{	-0.73	}$& $	63.32	^{+	11.00	}_{	-11.00	}$&$	-2.46	^{+	0.34	}_{	-0.34	}$&$	2.00	^{+	0.20	}_{	-0.20	}$(									10	-	1000	)&$(	5.50	^{+	0.40	}_{	-0.40	})\times 10^{	-6	}$(	10	-	1000	)&	GBM	&	10	\\
110213A	&	1.46	& $	-1.24	^{+	0.17	}_{	-0.17	}$& $	49.64	^{+	8.20	}_{	-8.20	}$&$	-2.08	^{+	0.05	}_{	-0.05	}$&$	17.70	^{+	0.50	}_{	-0.50	}$(									10	-	1000	)&$(	1.03	^{+	0.03	}_{	-0.03	})\times 10^{	-5	}$(	10	-	1000	)&	GBM	&	10	\\
110422A	&	1.77	& $	-0.65	^{+	0.036	}_{	-0.036	}$& $	152.00	^{+	3.04	}_{	-3.04	}$&$	-2.96	^{+	0.09	}_{	-0.12	}$&$							(1.01	^{+	0.08	}_{	-0.08	})\times 10^{	-5	}$(	20	-	2000	)&$(	8.56	^{+	0.01	}_{	-0.01	})\times 10^{	-5	}$(	20	-	2000	)&	K	&	13	\\
110503A	&	1.613	& $	-0.98	^{+	0.055	}_{	-0.049	}$& $	219.00	^{+	12.16	}_{	-11.55	}$&$	-2.75	^{+	0.12	}_{	-0.30	}$&$							(8.59	^{+	0.52	}_{	-0.52	})\times 10^{	-6	}$(	20	-	5000	)&$(	2.60	^{+	0.12	}_{	-0.12	})\times 10^{	-5	}$(	20	-	5000	)&	K	&	14	\\
110715A	&	0.82	& $	-1.23	^{+	0.055	}_{	-0.049	}$& $	120.00	^{+	7.29	}_{	-6.69	}$&$	-2.70	^{+	0.12	}_{	-0.30	}$&$							(7.06	^{+	0.39	}_{	0.39	})\times 10^{	-6	}$(	20	-	10000	)&$(	2.30	^{+	0.12	}_{	-0.12	})\times 10^{	-5	}$(	20	-	10000	)&	K	&	15 	\\
110731A	&	2.83	& $	-0.67	^{+	0.15	}_{	-0.15	}$& $	282.10	^{+	34.80	}_{	-34.80	}$&$	-2.64	^{+	0.42	}_{	-0.42	}$&$	20.90	^{+	0.50	}_{	-0.50	}$(									10	-	1000	)&$(	2.22	^{+	0.01	}_{	-0.01	})\times 10^{	-5	}$(	10	-	1000	)&	GBM	&	10	\\ \hline
\end{tabular}

\label{tab2}
}
}
\end{minipage}
\end{table*}
 
\begin{table*}
\begin{minipage}[t]{1 \hsize}
{\small
\caption{Intrinsic properties of 36 LGRBs}
\begin{tabular}{l|ccccccccc}
\hline
GRB&Type\footnote{I : Small-$ADCL$ GRBs  consistent with the Type I Fundamental Plane. \\II : Long tailed GRBs consistent with the Type II Fundamental Plane. \\
O : Outliers.} &$T_{90}$ &$T_{98}$ &$E_{\rm p}$\footnote{$E_{\rm p}=(1+z)E_{\rm p}^{\rm obs}$}
&$L_{\rm p, 2.752}^{\rm rest}$\footnote{Integrated between 1-10,000 keV in GRB rest frame. The flat $\Lambda$CDM universe with $H_0$ = 70 km s$^{-1}$ Mpc$^{-1}$ and $\Omega_{\rm M}$ = 0.30  is assumed. }
&$E_{\rm iso}$$^c$& $T_{\rm L}$& ADCL&\\
&&[seconds]&[seconds]&[keV]&[erg/sec]&[erg]&[seconds]& \\ \hline
970228	&	I	&	40.70	&	45.54	& $	194.93	^{+	64.41	}_{	-64.41	} $ & $(	7.34	^{+	1.21	}_{	-0.57	})\times 10^{	51	}$&$(	2.75	^{+	0.65	}_{	-0.38	})\times 10^{	52	}$&$	3.75	^{+	1.08	}_{	-0.60	}$&	0.14	\\
971214	&	I	&	7.05	&	9.62	& $	806.73	^{+	48.60	}_{	-63.18	} $ & $(	7.18	^{+	5.07	}_{	-2.34	})\times 10^{	52	}$&$(	3.97	^{+	2.34	}_{	-1.22	})\times 10^{	53	}$&$	5.53	^{+	5.09	}_{	-2.48	}$&	0.12	\\
990123	&	I	&	24.44	&	33.80	& $	1,333.80	^{+	49.92	}_{	-56.94	} $ & $(	2.27	^{+	0.10	}_{	-0.10	})\times 10^{	53	}$&$(	2.84	^{+	0.11	}_{	-0.10	})\times 10^{	54	}$&$	12.49	^{+	0.72	}_{	-0.70	}$&	0.09	\\
990506	&	I	&	56.32	&	60.94	& $	737.61	^{+	69.23	}_{	-87.86	} $ & $(	8.20	^{+	0.57	}_{	-0.57	})\times 10^{	52	}$&$(	9.80	^{+	0.64	}_{	-0.65	})\times 10^{	53	}$&$	11.96	^{+	1.14	}_{	-1.15	}$&	0.08	\\
990510	&	I	&	25.78	&	36.09	& $	538.20	^{+	25.14	}_{	-32.21	} $ & $(	3.33	^{+	0.85	}_{	-0.57	})\times 10^{	52	}$&$(	1.45	^{+	0.37	}_{	-0.25	})\times 10^{	53	}$&$	4.36	^{+	1.57	}_{	-1.06	}$&	0.14	\\
990705	&	I	&	16.91	&	20.35	& $	348.33	^{+	27.65	}_{	-27.65	} $ & $(	1.71	^{+	0.15	}_{	-0.13	})\times 10^{	52	}$&$(	2.49	^{+	0.43	}_{	-0.37	})\times 10^{	53	}$&$	14.56	^{+	2.85	}_{	-2.44	}$&	0.10	\\
991216	&	II	&	7.41	&	11.28	& $	1,083.73	^{+	37.37	}_{	-41.21	} $ & $(	2.19	^{+	0.11	}_{	-0.12	})\times 10^{	53	}$&$(	8.30	^{+	0.41	}_{	-0.44	})\times 10^{	53	}$&$	3.78	^{+	0.27	}_{	-0.29	}$&	0.20	\\
030329	&	I	&	13.53	&	18.08	& $	79.26	^{+	2.70	}_{	-2.51	} $ & $(	1.16	^{+	0.10	}_{	-0.09	})\times 10^{	51	}$&$(	1.55	^{+	0.07	}_{	-0.07	})\times 10^{	52	}$&$	13.41	^{+	1.27	}_{	-1.24	}$&	0.13	\\
050401	&	I	&	11.80	&	14.87	& $	458.25	^{+	70.20	}_{	-70.20	} $ & $(	4.42	^{+	0.91	}_{	-0.84	})\times 10^{	52	}$&$(	3.09	^{+	0.41	}_{	-0.41	})\times 10^{	53	}$&$	7.00	^{+	1.72	}_{	-1.62	}$&	0.15	\\
050525	&	I	&	4.50	&	6.10	& $	130.41	^{+	3.69	}_{	-3.69	} $ & $(	2.92	^{+	0.18	}_{	-0.17	})\times 10^{	51	}$&$(	2.32	^{+	0.12	}_{	-0.12	})\times 10^{	52	}$&$	7.95	^{+	0.64	}_{	-0.61	}$&	0.13	\\
050603	&	II	&	2.11	&	3.17	& $	1,313.28	^{+	332.43	}_{	-332.43	} $ & $(	1.36	^{+	0.22	}_{	-0.21	})\times 10^{	53	}$&$(	4.76	^{+	0.71	}_{	-0.68	})\times 10^{	53	}$&$	3.51	^{+	0.78	}_{	-0.74	}$&	0.23	\\
061007	&	I	&	24.80	&	27.82	& $	902.14	^{+	24.74	}_{	-26.12	} $ & $(	1.08	^{+	0.10	}_{	-0.08	})\times 10^{	53	}$&$(	9.82	^{+	0.41	}_{	-0.29	})\times 10^{	53	}$&$	9.13	^{+	0.97	}_{	-0.73	}$&	0.14	\\
070125	&	I	&	23.49	&	28.17	& $	934.75	^{+	100.63	}_{	-78.96	} $ & $(	1.07	^{+	0.10	}_{	-0.10	})\times 10^{	53	}$&$(	9.21	^{+	0.54	}_{	-0.45	})\times 10^{	53	}$&$	8.60	^{+	0.95	}_{	-0.90	}$&	0.10	\\
071003	&	O	&	9.44	&	13.67	& $	3,403.95	^{+	992.28	}_{	-992.28	} $ & $(	4.90	^{+	0.20	}_{	-0.59	})\times 10^{	52	}$&$(	4.09	^{+	0.30	}_{	-0.59	})\times 10^{	53	}$&$	8.35	^{+	0.71	}_{	-1.57	}$&	0.24	\\
071010B	&	II	&	7.13	&	9.93	& $	87.62	^{+	7.79	}_{	-13.63	} $ & $(	4.27	^{+	0.87	}_{	-0.58	})\times 10^{	51	}$&$(	2.35	^{+	0.67	}_{	-0.30	})\times 10^{	52	}$&$	5.50	^{+	1.92	}_{	-1.03	}$&	0.20	\\
080319B	&	O	&	24.38	&	38.49	& $	1,260.99	^{+	15.30	}_{	-16.48	} $ & $(	6.90	^{+	0.40	}_{	-0.40	})\times 10^{	52	}$&$(	1.33	^{+	0.02	}_{	-0.02	})\times 10^{	54	}$&$	19.26	^{+	1.16	}_{	-1.16	}$&	0.14	\\
080413B	&	II	&	2.29	&	3.99	& $	140.70	^{+	27.30	}_{	-16.80	} $ & $(	5.82	^{+	0.64	}_{	-0.50	})\times 10^{	51	}$&$(	1.85	^{+	0.19	}_{	-0.14	})\times 10^{	52	}$&$	3.18	^{+	0.47	}_{	-0.36	}$&	0.27	\\
080721	&	I	&	6.22	&	9.52	& $	1,746.97	^{+	146.71	}_{	-129.19	} $ & $(	3.34	^{+	0.32	}_{	-0.32	})\times 10^{	53	}$&$(	1.20	^{+	0.05	}_{	-0.05	})\times 10^{	54	}$&$	3.59	^{+	0.37	}_{	-0.37	}$&	0.16	\\
081121	&	I	&	4.65	&	5.30	& $	870.98	^{+	81.13	}_{	-68.32	} $ & $(	6.28	^{+	0.21	}_{	-0.14	})\times 10^{	52	}$&$(	2.48	^{+	0.30	}_{	-0.25	})\times 10^{	53	}$&$	3.95	^{+	0.50	}_{	-0.40	}$&	0.07	\\
081222	&	II	&	6.79	&	9.86	& $	567.39	^{+	59.57	}_{	-59.57	} $ & $(	6.73	^{+	1.04	}_{	-0.69	})\times 10^{	52	}$&$(	2.11	^{+	0.32	}_{	-0.21	})\times 10^{	53	}$&$	3.14	^{+	0.68	}_{	-0.45	}$&	0.30	\\
090323	&	I	&	12.52	&	13.96	& $	1,901.12	^{+	211.14	}_{	-202.80	} $ & $(	3.00	^{+	0.33	}_{	-0.32	})\times 10^{	53	}$&$(	3.85	^{+	0.32	}_{	-0.29	})\times 10^{	54	}$&$	12.84	^{+	1.76	}_{	-1.68	}$&	0.12	\\
090328	&	O	&	39.82	&	61.24	& $	1,281.34	^{+	116.31	}_{	-116.31	} $ & $(	7.52	^{+	1.03	}_{	-0.51	})\times 10^{	51	}$&$(	1.70	^{+	0.20	}_{	-0.06	})\times 10^{	53	}$&$	22.61	^{+	4.05	}_{	-1.72	}$&	0.20	\\
090424	&	II	&	31.17	&	40.17	& $	230.06	^{+	7.84	}_{	-7.84	} $ & $(	1.18	^{+	0.13	}_{	-0.10	})\times 10^{	52	}$&$(	4.67	^{+	0.29	}_{	-0.23	})\times 10^{	52	}$&$	3.97	^{+	0.51	}_{	-0.39	}$&	0.38	\\
090618	&	I	&	75.22	&	99.08	& $	229.61	^{+	11.03	}_{	-11.03	} $ & $(	1.19	^{+	0.08	}_{	-0.09	})\times 10^{	52	}$&$(	2.58	^{+	0.11	}_{	-0.11	})\times 10^{	53	}$&$	21.69	^{+	1.79	}_{	-1.93	}$&	0.10	\\
090902B	&	I	&	7.17	&	10.32	& $	2,235.31	^{+	38.66	}_{	-38.66	} $ & $(	4.83	^{+	0.05	}_{	-0.04	})\times 10^{	53	}$&$(	3.08	^{+	0.02	}_{	-0.02	})\times 10^{	54	}$&$	6.37	^{+	0.08	}_{	-0.07	}$&	0.09	\\
090926A	&	O	&	4.74	&	10.28	& $	895.52	^{+	30.38	}_{	-30.38	} $ & $(	3.84	^{+	0.16	}_{	-0.18	})\times 10^{	53	}$&$(	1.92	^{+	0.08	}_{	-0.07	})\times 10^{	54	}$&$	5.00	^{+	0.29	}_{	-0.30	}$&	0.26	\\
091003	&	O	&	14.64	&	22.57	& $	755.73	^{+	70.00	}_{	-70.00	} $ & $(	1.09	^{+	0.26	}_{	-0.04	})\times 10^{	52	}$&$(	9.16	^{+	2.00	}_{	-0.36	})\times 10^{	52	}$&$	8.40	^{+	2.71	}_{	-0.48	}$&	0.16	\\
091127	&	II	&	6.70	&	15.81	& $	54.09	^{+	3.41	}_{	-3.41	} $ & $(	1.99	^{+	0.08	}_{	-0.08	})\times 10^{	51	}$&$(	1.68	^{+	0.10	}_{	-0.04	})\times 10^{	52	}$&$	8.45	^{+	0.59	}_{	-0.37	}$&	0.32	\\
091208B	&	I	&	5.46	&	6.39	& $	257.26	^{+	82.52	}_{	-82.52	} $ & $(	5.49	^{+	2.06	}_{	-0.94	})\times 10^{	51	}$&$(	2.42	^{+	0.50	}_{	-0.27	})\times 10^{	52	}$&$	4.41	^{+	1.89	}_{	-0.90	}$&	0.07	\\
100414A	&	O	&	14.57	&	26.81	& $	1,356.39	^{+	38.36	}_{	-38.36	} $ & $(	5.85	^{+	0.77	}_{	-0.25	})\times 10^{	52	}$&$(	7.62	^{+	1.01	}_{	-0.35	})\times 10^{	53	}$&$	13.04	^{+	2.43	}_{	-0.82	}$&	0.21	\\
100906A	&	II	&	42.95	&	49.26	& $	310.33	^{+	68.72	}_{	-68.72	} $ & $(	3.15	^{+	0.37	}_{	-0.29	})\times 10^{	52	}$&$(	2.88	^{+	0.16	}_{	-0.13	})\times 10^{	53	}$&$	9.13	^{+	1.18	}_{	-0.93	}$&	0.22	\\
110213A	&	O	&	14.02	&	16.96	& $	122.11	^{+	20.17	}_{	-20.17	} $ & $(	1.67	^{+	0.15	}_{	-0.13	})\times 10^{	52	}$&$(	8.04	^{+	0.40	}_{	-0.37	})\times 10^{	52	}$&$	4.80	^{+	0.50	}_{	-0.44	}$&	0.09	\\
110422A	&	O	&	8.06	&	12.22	& $	421.04	^{+	8.42	}_{	-8.42	} $ & $(	1.58	^{+	0.12	}_{	-0.12	})\times 10^{	53	}$&$(	7.20	^{+	0.03	}_{	-0.03	})\times 10^{	53	}$&$	4.54	^{+	0.36	}_{	-0.36	}$&	0.10	\\
110503A	&	II	&	3.94	&	6.25	& $	572.25	^{+	31.77	}_{	-30.18	} $ & $(	6.07	^{+	0.33	}_{	-0.33	})\times 10^{	52	}$&$(	1.80	^{+	0.07	}_{	-0.07	})\times 10^{	53	}$&$	2.96	^{+	0.20	}_{	-0.20	}$&	0.29	\\
110715A	&	II	&	7.10	&	9.25	& $	218.40	^{+	13.28	}_{	-12.17	} $ & $(	1.09	^{+	0.09	}_{	0.04	})\times 10^{	52	}$&$(	4.81	^{+	0.20	}_{	-0.20	})\times 10^{	52	}$&$	4.42	^{+	0.40	}_{	-0.25	}$&	0.33	\\
110731A	&	II	&	2.52	&	9.42	& $	1,080.44	^{+	133.28	}_{	-133.28	} $ & $(	1.90	^{+	0.34	}_{	-0.18	})\times 10^{	53	}$&$(	4.55	^{+	0.44	}_{	-0.24	})\times 10^{	53	}$&$	2.39	^{+	0.49	}_{	-0.26	}$&	0.37	\\ \hline
 \end{tabular}
\label{tab:intrinsic}
}
\end{minipage}
\end{table*}

\end{document}